\documentclass[aip,apl,amsmath,amssymb,reprint]{revtex4-1}

\usepackage{graphicx}
\usepackage{dcolumn}
\usepackage{bm}
\usepackage{verbatim}
\usepackage[latin1]{inputenc}

\begin{document}

\preprint{APS/123-QED}

\title{Control of magnetic vortex wall chirality, polarity and position by a magnetic field}

\author{S. Hankemeier}
 \email{shankeme@physnet.uni-hamburg.de}
\author{A. Kobs}
\author{R. Frömter}
\author{H.P. Oepen}

\affiliation{
\textit{Universität Hamburg, Institut für Angewandte Physik, Jungiusstr. 11, 20355 Hamburg, Germany}\\
}
\date{\today}

\begin{abstract}
The seeding of vortex domain walls in V-shaped nanowires by a magnetic field has been investigated via simulations and Scanning Electron Microscopy with Polarization Analysis (SEMPA). It is found that the orientation of the magnetic field can be used to purposely tune the chirality, polarity and position of single vortex domain walls in soft magnetic nanowires.
\end{abstract}

\maketitle
A concept for potential application in memory devices \cite{Cowburn2007,Bohlens2008} is the micromagnetic vortex configuration, which is composed of a curling magnetization around a sharp core, where the magnetization is forced out-of-plane to minimize exchange energy. Stable magnetic vortices can be found as remanent state in circular microstructures as well as in nanowires as so called vortex domain walls in head-to-head or tail-to-tail domain arrangements \cite{Thiaville2007}. In microstructures it was found that the chirality, i.e. the sense of rotation of the in-plane magnetization, can be affected by magnetic structures in the vicinity \cite{Lua2008,Konoto2008,Hankemeier2009}. The ability to switch or set the chirality and the polarity, i.e. the magnetization of the core, is a field of intense research. Recent publications demonstrate the possibility to change the chirality in asymmetric single nanorings \cite{Jung2006,Nakatani2004,Klaui2001} and nanodisks \cite{Gaididei2008} by applying external fields. The polarity can be manipulated by a high frequency electrical current \cite{Yamada2007} or by a modulated magnetic field \cite{Hertel2007,Keavney2009,Weigand2009}. The reproducible manipulation of vortex wall properties is a necessary prerequisite for a storage concept based on vortex walls in combination with current induced domain wall movement, like in the racetrack memory device \cite{Parkin2008}.
\begin{figure}[b]
\includegraphics[width=0.95\linewidth]{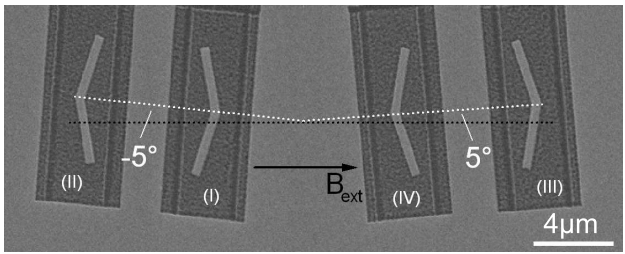}
\caption{\label{Fig1} Scanning Electron Microscope image of four V-shaped wires of 400~nm width. The axis of symmetry (bisection) is indicated by the white dotted line. The black arrow gives the direction of the external field used for seeding the domain walls.}
\end{figure}
\begin{figure}
\includegraphics[width=0.95\linewidth]{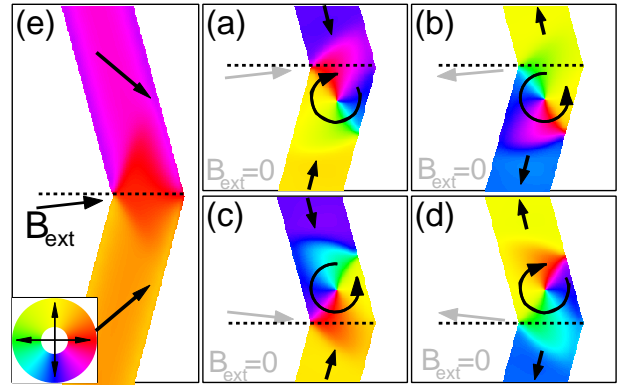}
\caption{\label{Fig2} (Color online) Simulation of the magnetization of V-shaped wires. The magnetization orientation is indicated by the black arrows and color coded according to the color wheel. The symmetry axis of the wire (bisection) is illustrated by the black dotted lines. In (a)-(d) the remanent states after pretreatment in an external field in the direction of the gray arrows is shown. In (e) the magnetic configuration during an applied external field of $B_{ext}=60$~mT is presented.}
\end{figure}
\begin{figure*}[t]
\includegraphics[width=0.95\linewidth]{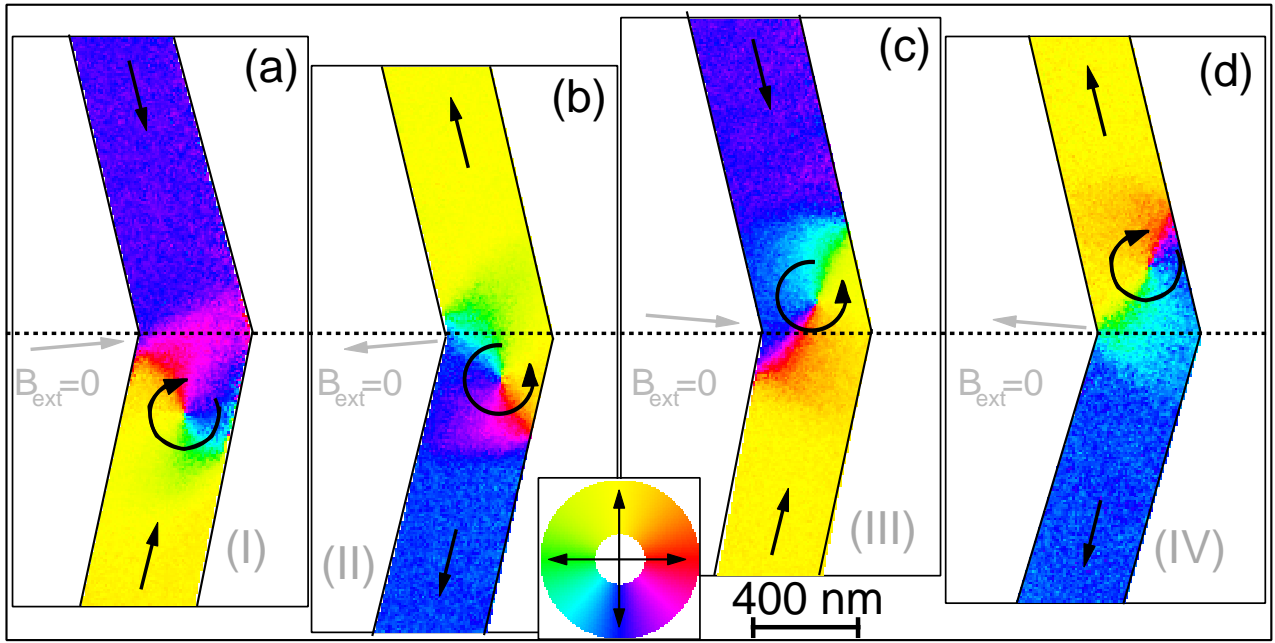}
\caption{\label{Fig3} (Color online) Remanent SEMPA images of V-shaped wires after field application in the indicated direction. The labels (I)-(IV) refer to the four arrangements shown in Fig.~\ref{Fig1}. The images have been rotated and the bisections of the wires are aligned as indicated by the black dotted line. The orientation of magnetization is indicated by the black arrows and color coded according to the color wheel. }
\end{figure*}
A common approach \cite{Taniyama1999,Taniyama2000} to introduce a domain wall into a V-shaped nanowire is to apply a magnetic field along the line of intersection of the two arms (see Fig.~\ref{Fig1}). Depending on the sign of the external field, a head to head or a tail to tail wall nucleates. While the magnetization orientation in the domains is well understood, it is still an open question what determines the domain wall properties. More precise, the reason for the appearance of asymmetric domain wall configurations, like the vortex wall, in a structure with mirror symmetry has to be found. When vortex walls are generated, the core is in general slightly shifted out of the symmetry axis into one arm of the V-shaped wire \cite{Brownlie2006}, in contrast to transverse walls which comply with the mirror symmetry. In this paper it is demonstrated that the orientation of the seeding field is the control that determines the location and chirality of the vortex wall. A further side effect is that the polarity of the vortex is also affected and thus becomes experimentally accessible. To investigate that issue, we have performed simulations by OOMMF \cite{OOMMF} and experiments via Scanning Electron Microscopy with Polarization Analysis (SEMPA) \cite{Froemter2009}.
First we have investigated the seeding of vortex walls via micromagnetic simulations. We have simulated four generic cases for the field orientation with respect to the wire. The magnetic field was tilted by $\alpha=\pm5^\circ$ out of the symmetry axis, which is the center line through the wire within the bent (see Fig.~\ref{Fig1}). The field strength was $B_{ext}=60$~mT in the direction indicated by the gray arrows. The results of the simulation in remanence after switching off the field are plotted in Figs.~\ref{Fig2}(a)-(d). The input parameters for the simulations were $M_{S}=1,430~\frac{\text{kA}}{\text{m}}$, $A=35~\frac{\text{pJ}}{\text{m}}$, cell-size = 4~nm~x~4~nm~x~18~nm, wire width = 400~nm, height = 18~nm and a bending angle of 150$^\circ$. These parameters favor the vortex wall over the transverse wall configuration \cite{Hankemeier2009b}.
Fig.~\ref{Fig2}(e) shows the magnetic domain structure in the wire when a field of $B_{ext}=60$~mT is applied at $\alpha=-5^\circ$ (black arrow) with respect to the bisection of the wire. This state transforms into the configuration shown in Fig.~\ref{Fig2}(a) when the field is switched off. This domain pattern reveals a head-to-head vortex wall. The vortex core settles in the lower arm of the V-shaped wire, while the chirality of the vortex is clockwise (CW) oriented. The core is located 212~nm into the arm of the wire and is in addition laterally shifted 20~nm towards the outer edge. Occupying this position, the energy of the domain wall is minimized as the curling magnetization of the vortex can easily follow the outer edge and has more space to comply with the abrupt change of the edge-direction at the inner bend. Reversing the magnetic field ($B_{ext}=-60$~mT) creates in the relaxed state at zero field a tail-to-tail wall. The vortex core is again placed in the lower arm, while the chirality is switched from CW to counter-clockwise (CCW) (see Fig.~\ref{Fig2}(b)). In Figs.~\ref{Fig2}(c)(d) the seeding field was applied at $\alpha=+5^\circ$ with $B_{ext}=\pm60$~mT out of the high symmetry direction. The configuration in zero field shows a head-to-head / tail-to tail wall, respectively. The vortex, however, nucleates in the upper part of the wire for both field directions. The chirality again depends on the sign of the applied field, a CCW/CW orientation is found for $\pm60$~mT (Figs.~\ref{Fig2}(c)(d)). To summarize the results: The orientation of the external seeding field determines which of the four micromagnetic configurations (Figs.~\ref{Fig2}(a)-(d)) occurs. During application of the external field the energetical degeneration of the four states is lifted and one configuration is favored: The vortex core shifts into the arm that has the smallest angle to the field direction, while the chirality is governed by the magnetization of the field-aligned apex of the wire. Hence, the chirality of the vortex is reversed on change of the sign of the external field at a fixed angle.
To prove the results of the simulations experimentally, V-shaped wires of 400~nm width and a bending angle of 150$^\circ$ have been carved out of 18~nm thick, soft magnetic CoFeSi film via Focused Ion Beam (FIB) milling (Fig.~\ref{Fig1}). The CoFeSi film with a composition ratio of 39\% 54\% 7\% (atomic~\%) has been electron-beam evaporated on a silicon single crystal substrate. The magnetic properties of the film \cite{Hankemeier2009b} are very close to the values used in the simulation. Wires were created with tilt angles of $\pm5^\circ$ ((I) and (III) in Fig.~\ref{Fig1}) and $\pm175^\circ$ ((II) and (IV) in Fig.~\ref{Fig1}) to the axis, along which the field is applied in the experiment. The former/latter two wires represent the arrangement of the simulation displayed in Figs.~\ref{Fig2}(a)(c) / Figs.~\ref{Fig2}(b)(d), respectively. This arrangement of the wires allows to investigate all four generic cases of field orientation simultaneously.
After applying a magnetic field of $B_{ext}=60$~mT in the direction indicated by the black arrow in Fig.~\ref{Fig1}, the domain pattern of the ensemble of wires is studied.
SEMPA images of the four generic arrangements are shown in Figs.~\ref{Fig3}(a)-(d). Comparing Fig.~\ref{Fig2} and Fig.~\ref{Fig3} it is obvious that the experiment confirms the simulations. A peculiar difference of experiment and simulation is the exact location of the vortex. While the simulation gives a constant distance of the core to the bisection of 212~nm, the experiment shows a variation of the distance ranging from 125~nm to 310~nm. The lateral displacement towards the edge varies from 15~nm to 45~nm, while the simulation gives a value of 20~nm. We attribute these deviations to pinning by microscopic magnetic irregularities, which might be due to the grain structure of the magnetic film.
\begin{figure}[t]
\includegraphics[width=0.95\linewidth]{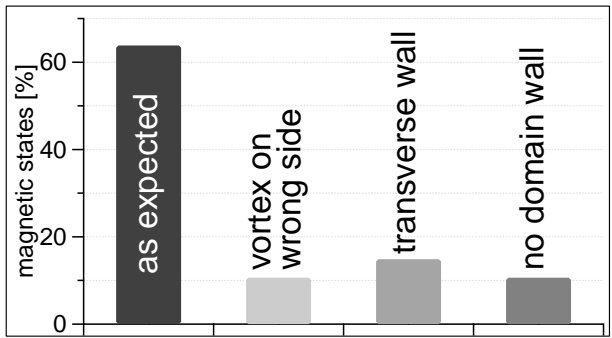}
\caption{\label{Fig4} Bar chart of the success rate for the nucleation of a vortex wall configuration as shown in Fig.~\ref{Fig2} and Fig.~\ref{Fig3}.}
\end{figure}
Cyclic reversal of the external field, 47 magnetization processes in an array of wires have been performed and analyzed. Fig.~\ref{Fig4} gives the frequency distribution of the obtained domain wall configurations observed in the experement. Beside vortex walls, we also found transverse walls (16\%) and in 11\% of the magnetization processes no domain wall was found. A 63\% majority of all examined magnetization processes gives vortex patterns that agree with the simulations (Fig.~\ref{Fig2}), which is a reasonable success rate compared to similar experiments \cite{Konoto2008,Backes2007}. A wrong vortex configuration is found with a probability of 11\%. For technical applications a further optimization of parameters is required.
The polarization of the core is another important magnetic feature of the vortex, as it describes the orientation of the magnetization within the center of the core (up or down). The simulations reveal that a magnetic field component of 70~mT perpendicular to the film plane, applied during vortex core nucleation, causes the polarization of the vortex core to follow the out-of-plane field. This implies that the polarity can be set on purpose, needing less than 4\% of the saturation magnetization of 1.8~T that aligns all the magnetic moments completely out of plane.

In summary, we have demonstrated that the chirality and the position of a vortex domain wall in a V-shaped wire can be controlled via the orientation of the seeding magnetic field. The possibility to purposely control both the chirality and the polarity of a vortex domain wall gives more flexibility in future concepts of vortex based memory devices: A V-shaped injection wire can be used to define a single vortex configuration which acts as a four state bit element and can be moved into a memory array utilizing the spin torque effect, in analogy to the working principle of the racetrack memory \cite{Parkin2008}.
\begin{acknowledgments}
Financial support by DFG via SFB 668 is gratefully
acknowledged.
\end{acknowledgments}
%
\end{document}